\documentclass[journal]{IEEEtran}

\usepackage{amsmath}
\usepackage{wasysym}
\usepackage{graphicx}
\usepackage{subfig}
\usepackage{cite}

\begin{document}

\title{High-Temperature Superconducting Spiral Resonator for Metamaterial Applications}

\author{Behnood~G.~Ghamsari, John Abrahams, Stephen Remillard,  and Steven M. Anlage

\thanks{
This work is, in part, supported by the DOE grant number DESC0004950 and ONR/AppEl, Task D10, through grant
number N000140911190.

B.G.G, J.A, and S.M.A are with the Center for Nanophysics and Advanced Materials, Department of Physics, University of Maryland, College Park, MD, 20742-4111 USA.
(e-mail: ghamsari@umd.edu, anlage@umd.edu)

S.R is with the Department of Physics, Hope College, Holland, MI, 49422-9000 USA.}}

\maketitle

\begin{abstract}
This work studies high-temperature superconducting spiral resonators as a viable candidate for realization of RF/microwave metamaterial atoms.
The theory of superconducting spiral resonators will be discussed in detail, including the mechanism of resonance, the origin of higher order modes, the analytical framework for their determination, the effects of coupling scheme, and the dependence of the resonance quality factor and insertion loss on the parity of the mode. All the aforementioned models are compared with the experimental data from a micro-fabricated YBa$_2$Cu$_3$O$_{7-\delta}$ (YBCO) spiral resonator. Moreover, the evolution of the resonance characteristics for the fundamental mode with variation of the operating temperature and applied RF power is experimentally examined, and its implications for metamaterial applications are addressed.
\end{abstract}

\begin{IEEEkeywords}
High-Temperature Superconducting Spiral Resonators, RF/Microwave Metamaterial, Magnetic Metamaterial Atoms.
\end{IEEEkeywords}

\section{Introduction}
\PARstart{S}UPERCONDUCTING resonators have been long used in science and technology due to their low-losses and high quality factors.
Recently, these structures have been seriously sought to be utilized in metamaterial applications, not only because of their low-loss, but also for their deep sub-wavelength dimensions and tunability\cite{Anlage}.

Various metamaterial schemes have been recently proposed and implemented based on both fully superconducting devices and in combination with normal conductors\cite{Ricci1,Ricci2,Ricci3,Salehi,Fedotov,Kurter1}.
In particular, superconducting spiral resonators (SSRs) have been introduced as compact self-resonating structures at RF frequencies, whose physical dimensions are several orders of magnitude smaller than the operating wavelength\cite{Kurter2,Kurter3}. Therefore, SSRs demonstrate great potential for constructing one-, two-, and three-dimensional metamaterials.

Inasmuch as the majority of the studies on SSRs have focused on low-temperature superconductors (LTS), such as Nb spirals, this work concerns high-temperature superconducting (HTS) spiral resonators, which generally exhibit greater tunability and higher nonlinearity than their LTS analogues.

Our device is made of a 300nm-thick YBCO film on a sapphire substrate and is patterned by standard photolithography techniques into a 6mm-diameter spiral consisting of 40 turns of 10$\mu m$-wide lines with 10$\mu m$ spacing.
As Fig. \ref{Coupling} depicts, the device is measured by two loop antennas: one approaching the device from the top and one from the bottom.
More details about the patterning process and the measurement setup can be found elsewhere\cite{Ghamsari}.

In the next section, the theory of SSRs is thoroughly discussed. At each stage, the results of our analysis is compared with the experimental data measured from the YBCO spiral. Section III, examines the effects of the operating temperature and excitation power on the resonance characteristics including the resonance frequency, quality factor, and insertion less. Finally the implications of these effects for metamaterial applications are addressed, and the conclusion follows.

\section{Theory of Superconducting Spiral Resonators}

In the absence of the spiral resonator, the impedance matrix of the coupled exciting loop antennas reads
\begin{equation} \label{ZLoops}
[Z]= \begin{bmatrix} j\omega L_1 & j\omega M_{12} \\
j\omega M_{12} & j\omega L_2 \end{bmatrix},
\end{equation}
where $L_1$ and $L_2$ are the self inductances of the loop antennas, $M_{12}$ is their mutual inductance, and $\omega$ is the angular frequency of the signal. Accordingly,
\begin{equation} \label{SLoops}
S_{21} = \frac{2j\omega M_{12}Z_0}{Z_0^2+j\omega(L_1+L_2)Z_0 + (M_{12}^2-L_1L_2)\omega^2},
\end{equation}
where $Z_0$ is the characteristic impedance of the transmission lines feeding the antennas.
The spiral resonator may be modeled, to first approximation, by a parallel RLC resonator that is magnetically coupled to the loops, as shown by Fig. \ref{Coupling}. Therefore,
the loaded impedance matrix is
\begin{equation} \label{SPLoops}
[Z]= \begin{bmatrix} j\omega L_1+\frac{M_{10}^2\omega^2}{Z_{RLC}} & j\omega M_{12}+\frac{M_{10}M_{20}\omega^2}{Z_{RLC}}\\\\
j\omega M_{12}+\frac{M_{10}M_{20}\omega^2}{Z_{RLC}} & j\omega L_2 +\frac{M_{20}^2\omega^2}{Z_{RLC}}\end{bmatrix},
\end{equation}
where {$M_{10}$ and $M_{20}$ are the mutual inductances between the loop antennas and the spiral resonator and $Z_{RLC}=j\omega L+R/(1+j\omega RC)$. The scattering matrix can be straightforwardly evaluated from $S = ([Z]+[Z_0])^{-1}([Z]+[Z_0])$\cite{Pozar}.\\
\begin{figure}
\centering{\subfloat[][]{\includegraphics[width =3in]{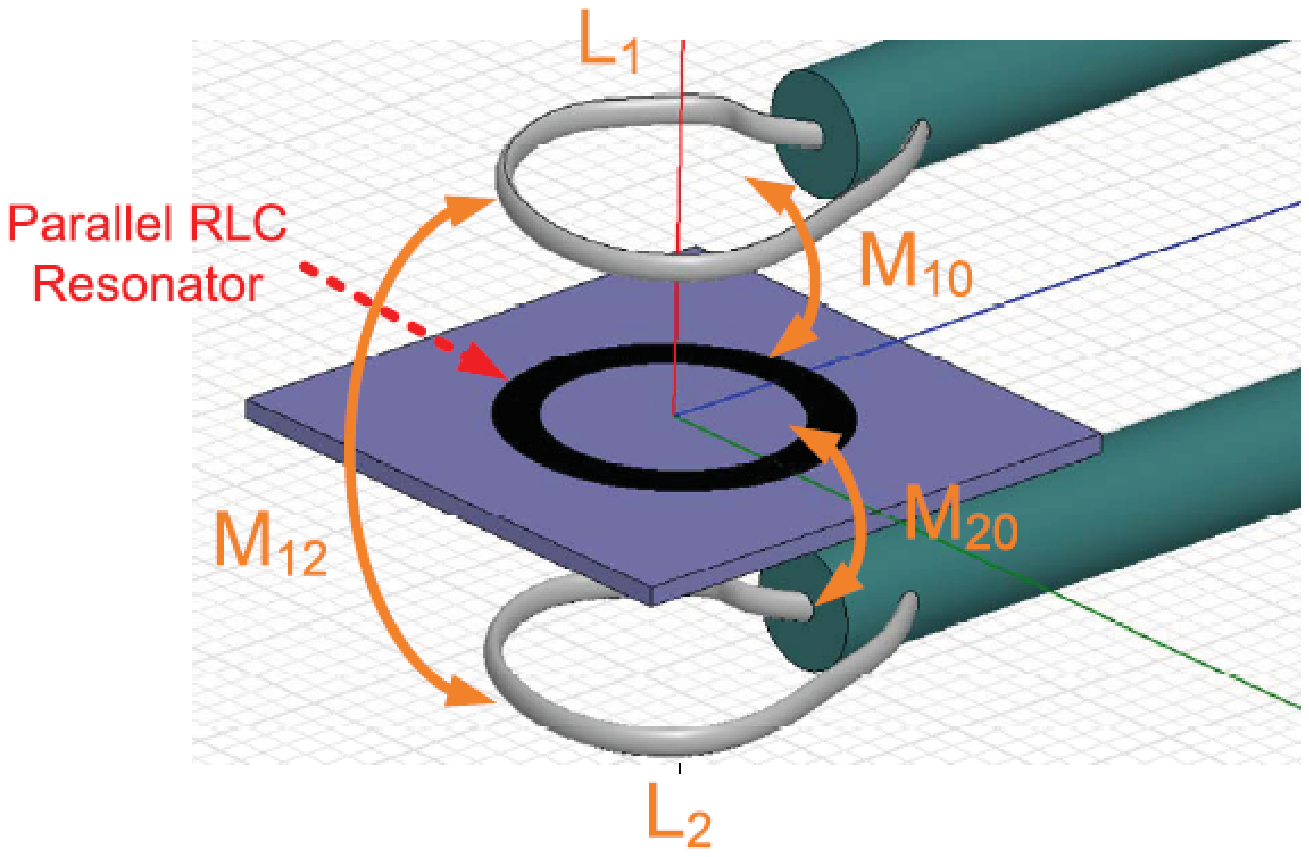}
\label{Coupling}}
\hspace{8pt}%
\subfloat[][]{\includegraphics[width=3.5in]{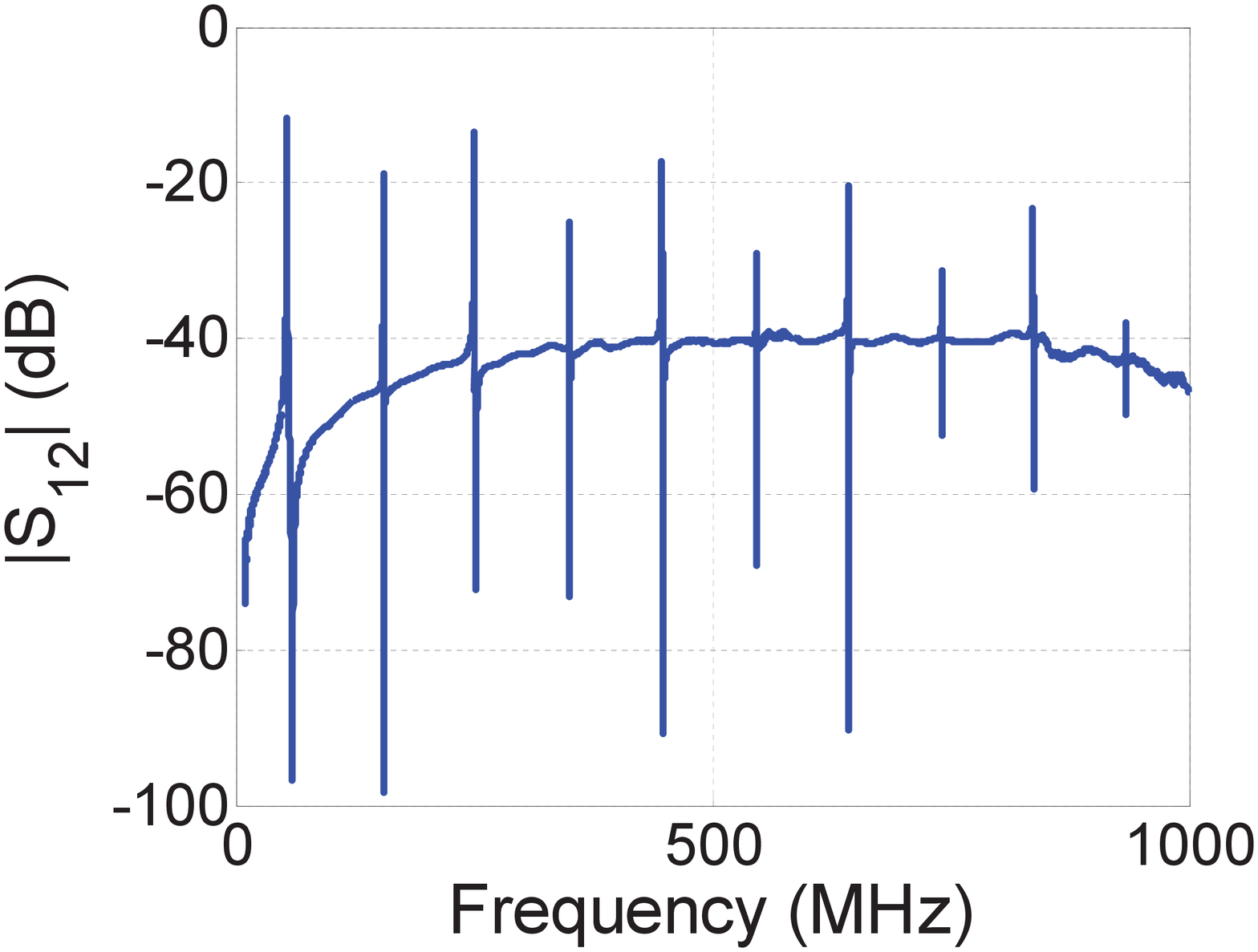}
\label{Spectrum}}\\
}
\captionsetup{font=small}  \caption{(a) The schematic of the coupling between the loop antennas and the spiral resonator. (b) The transmission spectrum of the YBCO spiral resonator including ten resonances below 1GHz. The fundamental resonance frequency is 53.6MHz with a Q as high as 1400 at 10K.}\label{Spiral}
\label{sp}
\end{figure}
As Figure \ref{model} illustrates, the analytical model of (\ref{SPLoops}) is in excellent agreement with experiment.
The enhanced transmission peak is due to the constructive interference between the loops direct coupling through $M_{12}$ and the indirect coupling mediated by the spiral resonator through $M_{10}$ and $M_{20}$. The dip in transmission, which immediately follows, is due to the destructive interference between the two coupling channels that essentially cancel each other.
Nevertheless, the spiral resonator also possesses high order resonant modes, as shown in Fig. \ref{Spectrum}. In order to account for the complete spectrum of the spiral resonator, one should consider the device as an inductively coupled transmission line resonator. Therefore, the resonance condition reads as $\Im m\{Z_{21}(\omega)\}=0$. Ignoring the losses in the superconducting resonator and replacing $Z_{RLC}$ by the impedance seen at the input of the transmission line, one gets the following secular equation for finding the resonances:
\begin{equation}\label{ModeCondition}
\tan(\beta\ell-\phi) = \frac{Z_cM_{12}}{\omega M_{10}M_{20}},
\end{equation}
where $\beta$ and $Z_c$ respectively are the propagation constant and the characteristic impedance of the transmission line, $\phi$ is the equivalent termination of the transmission line, which is ideally open circuited, and $\ell$ is the effective length of the Archimedean spiral. Most commonly, transmission line resonators are capacitively coupled, leading to a qualitatively different secular equation \cite{Pozar}
\begin{equation}\label{ModeCondition2}
\tan(\beta\ell-\phi) = -\omega CZ_c.
\end{equation}
\begin{figure}
\begin{center}
\includegraphics[width= 3 in]{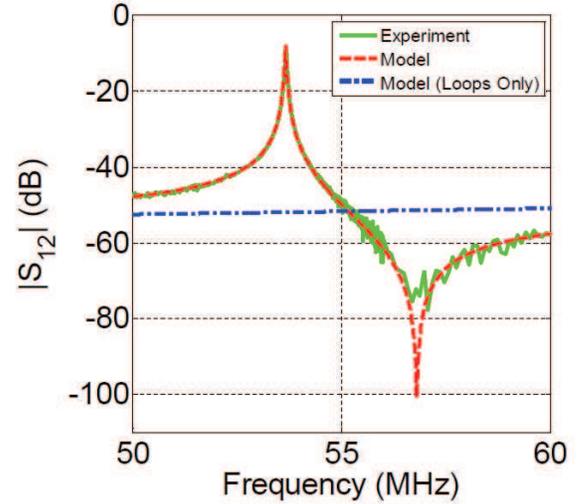}
\captionsetup{font=small} \caption{Comparison between the theory and experiment for the fundamental mode of the spiral resonator.
The solid line curve is the experimental data for a YBCO spiral at 10K, the dash-doted curve is the background coupling between the loops according to equation (\ref{SLoops}), and the dashed curve is the analytical model of equation (\ref{SPLoops}) with the following
parameters: $L_1=L_2=12nH$, $M_{10}=M_{20}=3.2nH$, $M_{12}=190pH$, $L=500nH$, $C=17.6pF$, $R=400k\Omega$, and $Z_0=50\Omega$
Note that $L_1$, $L_2$, and $M_{12}$ are determined by the $|S_{21}|$ data above $T_c$ of the YBCO.}\label{model}
\end{center}
\end{figure}
Figure \ref{Modes} illustrates the solutions to equation (\ref{ModeCondition}), with the parameters listed in Fig. \ref{model}, as well as the experimentally measured resonance frequencies for our YBCO spiral resonator.
The fundamental mode is nearly at $(\beta\ell-\phi)\approx\pi/2$; however, higher order modes gradually deviate from $(\beta_n\ell-\phi)\simeq n\pi/2$ and approach $(\beta_n\ell-\phi)\simeq n\pi$ for large $n$.
For comparison, the resonances of a capacitively coupled resonator of similar character are also shown in the same figure, where the fundamental mode is far from $(\beta\ell-\phi)\approx\pi/2$; however, the higher order modes progressively  approach $(\beta_n\ell-\phi)\simeq n\pi/2$ for large $n$.\\
\begin{figure}
\begin{center}
\includegraphics[width= 3.5 in]{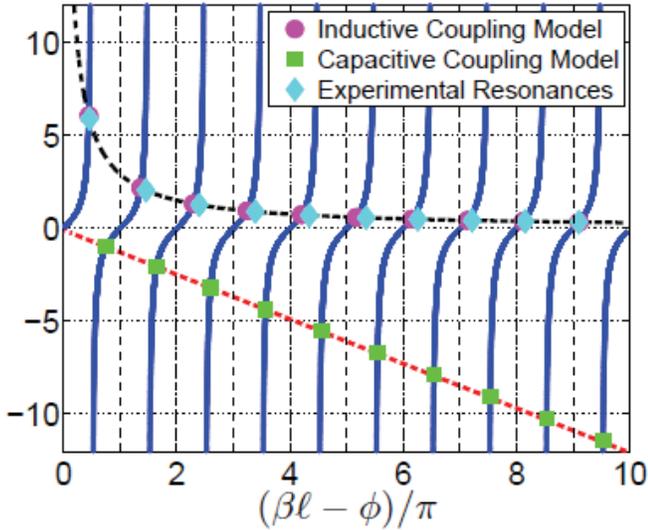}
\captionsetup{font=small} \caption{Plot of the right hand and left hand sides of equations (\ref{ModeCondition}) and (\ref{ModeCondition2}).
Intersections indicate solutions to the mode conditions (\ref{ModeCondition}) and (\ref{ModeCondition2}). The red circles and green squares correspond to the resonance modes for the inductive and capacitive coupling models, respectively, and
the blue diamonds are the experimental resonances of the YBCO spiral resonator.
}\label{Modes}
\end{center}
\end{figure}
Interestingly, measurement reveals that the even order resonances of the spiral possess considerably higher quality factors and insertion losses compared to the adjacent odd order modes. Referring to the earlier studies of laser scanning microscopy on Nb spiral resonators\cite{Zhuravel}, the spatial distribution of RF current over a spiral resonator comprises alternatively clockwise and counterclockwise current bands, whose total number equals the mode order, i.e. one current band in the fundamental mode, two counter rotating current bands for the second resonance and so forth. Obviously, even order modes have equal number of clockwise and counterclockwise current bands; thus, their net magnetic moment is nearly zero. In contrast, odd order modes, due to unequal number of counter-circulating current bands, always carry a significant magnetic moment, which, in turn, strongly couples to the exciting loop antennas. Earlier calculation of the mutual inductance between the spiral and the loops for different modes of Nb spirals confirms this interpretation\cite{Kurter3}.
A stronger coupling subsequently leads to a higher transmission and a lower quality factor for the odd order modes.
The highest mutual inductance is evidently associated with the fundamental mode, which experimentally always has the lowest quality factor and the lowest insertion loss.
The coupling between the spiral resonator and the antennas is a strong function of their separation\cite{Kurter2}, therefore, there is always a trade-off between a high quality factor and a high transmission depending on the height of the exiting loops.

\section{Temperature and Power Dependence}

Figure \ref{TDependence} shows that the fundamental mode, at a low excitation power, is not severely distorted over a wide range of temperatures; however, at elevated temperatures, typically higher than $0.7T{_c}$, the disturbance is pronounced and even the shape of the resonance deviates from a Lorentzian peak\cite{Zaitsev}. Similarly, as Fig. \ref{PDependence} illustrates, high RF powers can drastically disturb the resonance, especially at high temperatures.

\begin{figure}
\begin{center}
\includegraphics[width= 3.5 in]{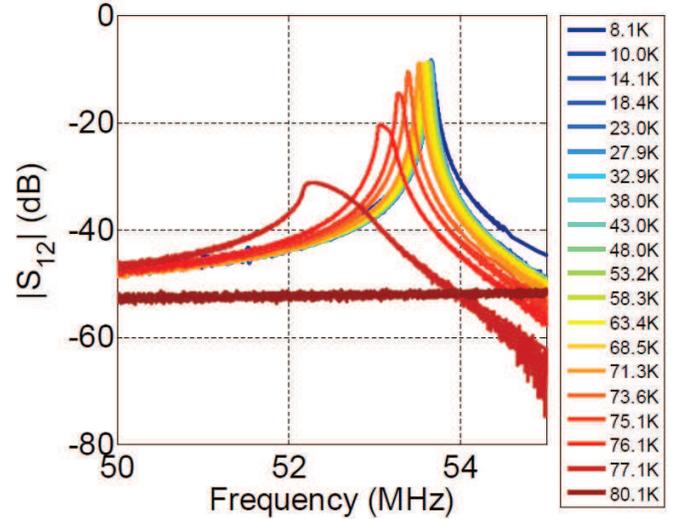}
\captionsetup{font=small} \caption{Experimental transmission spectrum of the fundamental mode of the YBCO spiral resonator with varying operating temperatures at a fixed -12dBm input power.}\label{TDependence}
\end{center}
\end{figure}

\begin{figure}
\begin{center}
\includegraphics[width= 3.5 in]{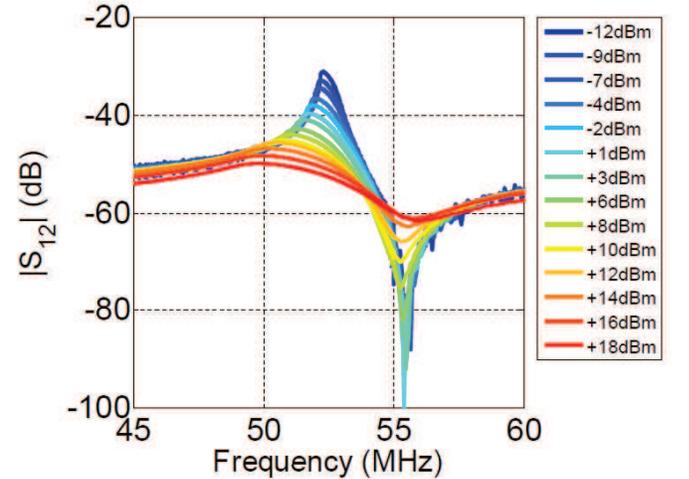}
\captionsetup{font=small} \caption{Variation of the fundamental mode of the YBCO spiral resonator as a function of input RF power at 77K.}\label{PDependence}
\end{center}
\end{figure}

In principle, an increase in the temperature or RF power suppresses the superconducting order parameter and reduces the superfluid density, and at the same time raises the normal conductivity of the sample by generating quasiparticles. The reduction of the superfluid density yields a higher kinetic inductance, and, therefore, shifts the resonance to lower frequencies. On the other hand, the rise in the sample normal conductivity translates to more dissipation, which lowers the quality factor and increases the insertion loss.

\begin{figure}
\begin{center}
\captionsetup{font=small} \includegraphics[width= 3.5 in]{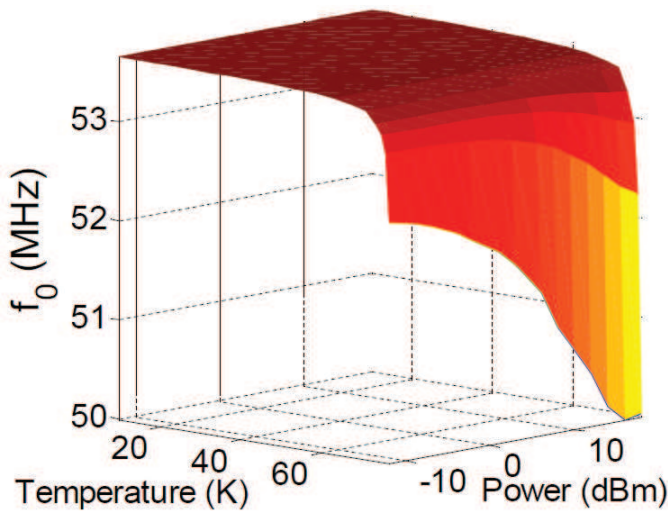}
\caption{Experimental variation of the fundamental resonance frequency of the YBCO spiral resonator with temperature and RF power.}\label{F}
\end{center}
\end{figure}

All three of these effects are clearly evident from the experimental data shown in Fig. \ref{F} to Fig. \ref{IL}. The plateaus in the figures, however, define the range of temperature and input RF power where the spiral resonator is robust, and even though its resonance characteristics slightly change with the stimuli, generally they remain very close to the low-temperature values. At this regime, these slight changes have almost a linear dependence on the strength of the excitation; thus, the plateau hosts a suitable quiescent point for tunable RF/microwave devices and metamaterials. The extent of the plateaus for the spiral resonator can be partly controlled by adjusting the coupling strength between the loop antennas and the device through the height of the loop antennas. Operating conditions similar to the plateaus of Fig. \ref{F} to Fig. \ref{IL} have been also used for realizing optically tunable superconducting microwave devices\cite{Atikian}.

At high temperatures and RF powers the resonance characteristics drastically change in a nonlinear fashion.
This can be very useful for switching the superconducting meta-atoms into a high-loss state\cite{Zhuravel,Kurter4}.
At temperatures too close to $T_c$ the resonator looses many of its useful features, such as a high quality factor, and probably is less attractive for device applications. However, this regime is interesting for probing the intrinsic properties of the superconducting film\cite{Tai}.
Nevertheless, at low and moderate temperatures, with a high RF power one can simultaneously benefit from the nonlinear behavior of the resonator along side the usual desired features like a high Q. This regime is especially well-suited for implementation of parametric processes such as parametric amplification and potentially RF metamaterials with gain.

\begin{figure}
\begin{center}
\includegraphics[width= 3.5 in]{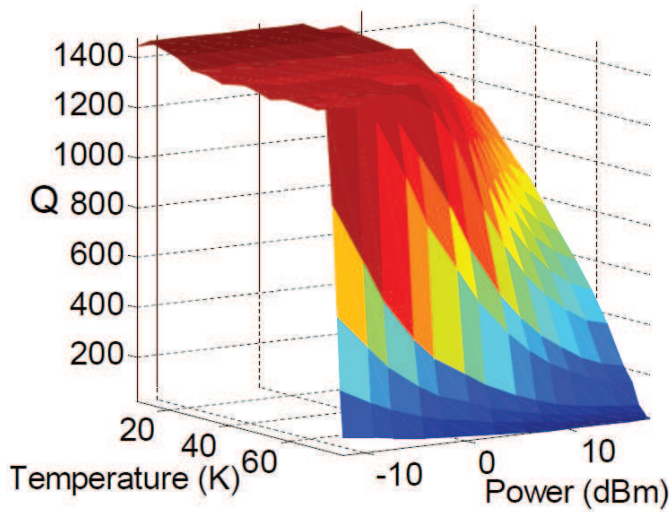}
\captionsetup{font=small} \caption{Experimental variation of the quality factor for the fundamental mode of the YBCO spiral resonator with increasing temperature and RF power.}\label{Q}
\end{center}

\end{figure}
\begin{figure}
\begin{center}
\includegraphics[width= 3.5 in]{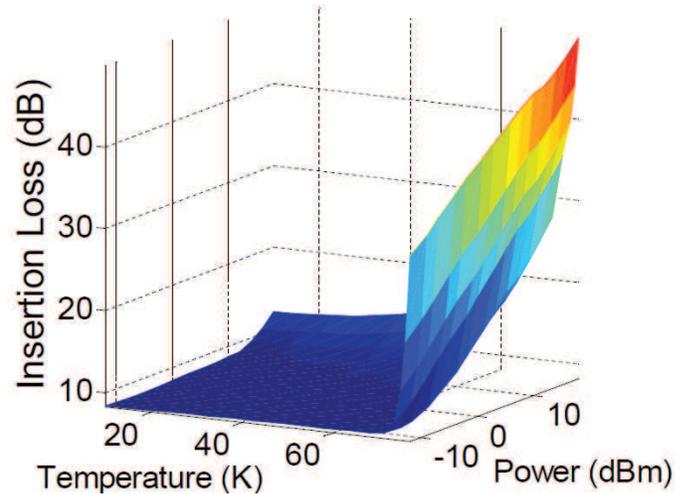}
\captionsetup{font=small} \caption{Experimental variation of the insertion loss for the fundamental mode of the YBCO spiral resonator with high temperatures and RF powers.}\label{IL}
\end{center}
\end{figure}

\section{Conclusion}
We presented the theory of superconducting spiral resonators in detail, including the mechanisms and conditions of resonance as well as the effect of the coupling scheme. We showed that our model is in excellent agreement with the experiment. Moreover, we experimentally studied the temperature and power dependence of the YBCO spiral resonator, identified different regimes of interest, and highlighted the potential of integrating parametric processes, such as parametric amplification, with metamaterial functions.

\end{document}